\documentstyle[editedvolume,numreferences,epsfig]{crckapb} 

\begin{opening}
\title{Physics Objectives for Future Studies\protect\\
of the Spin Structure of the Nucleon}
\author{Wolf-Dieter Nowak}
\institute{DESY Zeuthen, Platanenallee 6, D-15738 Zeuthen, Germany \\
e-mail: Wolf-Dieter.Nowak@desy.de}
\end{opening}

\begin{document}

\begin{abstract}
Physics perspectives are shown for future experiments in electron or 
positron scattering on nucleons, towards a deep and comprehensive 
understanding of the angular momentum structure of the nucleon in the 
context of Quantum Chromodynamics. Measurements of Generalised 
Parton Distributions in exclusive reactions and precise determinations
of forward Parton Distributions in semi-inclusive deep inelastic 
scattering are identified as major physics topics. Requirements are 
discussed for a next generation of high-luminosity fixed-target experiments
in the energy range 30-200 GeV.\footnote{Invited talk at NATO `Advanced Spin
Physics Workshop', June 30-July 3, 2002, Nor-Hamberd, Armenia}
\end{abstract}

\section{Introduction}
Charged leptons have been used for more than two decades as a
very powerful tool for studying the {\it momentum structure}
of the nucleon, in a wide variety of experimental approaches.
More recently, high-energy polarised beams 
and high-density polarised targets have become accessible and have
proven to be indispensable tools in studying the {\it angular 
momentum structure} of the nucleon. \\

At large enough $Q^2$, the four-momentum-transfer squared of the photon 
mediating the lepton-nucleon interaction, short-range phenomena 
(`hard' photon-parton interactions) are successfully described by 
perturbative Quantum Chromodynamics (QCD). In contrast, long-range (`soft') 
phenomena and especially parton correlations in hadrons, 
over a broad range in $Q^2$, are still lacking a satisfactory theoretical 
description. Here it may be expected that contemporary 
theoretical developments  will at some point turn into a calculable 
field theoretical description of hadronic structures.  \\

The spin of the nucleon as a whole is $\frac{\hbar}{2}$, irrespectively of
the resolving power $Q^2$ of the virtual photon. Due to angular 
momentum conservation in QCD the individual contributions of the parton's 
spins and orbital angular momenta always add up to this value, although
they vary considerably with $Q^2$. Precise measurements at high and
especially moderate $Q^2$ are required to be able
to reliably `bridge' into the `critical' region of $Q^2 \leq 0.5$ GeV$^2$
to eventually test theoretical descriptions of soft phenomena.  \\

In this paper, based on recent theoretical developments in the
field, an experimentalist's perspective is given on physics
prospects for possible future electron-nucleon fixed-target experiments 
in the center-of-mass energy range of up to a few tens of GeV$^2$, with high 
resolution, high luminosity and polarised beams and/or targets.  \\

\section{Overview of the Relevant Quantities}
%
The cross section of the inclusive reaction 
$e \, N \, \rightarrow \, e \, X$ is not yet exactly calculable from theory.
Instead it is presently parameterised by non-perturbative `structure 
functions' which in turn, in the framework of the quark-gluon picture of 
the nucleon, are expressed in terms of Parton Distribution Functions
(PDFs). These functions have proven to be very useful in the description 
of the momentum and spin structure of partons. However, they
do not yet include the parton's {\it orbital} angular momenta which are 
considered to be essential among the various components that eventually
make up the half-integer spin of the nucleon. This striking deficiency 
is cured in the formalism of `Generalized Parton Distributions' 
(GPDs)~\cite{Mueller,Ji,Radyushkin,Bluemlein} which reached the level of 
practical applications only recently. This theoretical framework is
capable of simultaneously treating several types of processes ranging 
from inclusive to hard exclusive scattering. Exclusive lepton-nucleon
scattering is `non-forward' in nature since the photon initiating the 
process is virtual and the final state particle is usually real, forcing a 
small
but finite momentum transfer to the target nucleon. While it is very appealing
that GPDs embody forward (`ordinary') PDFs as well as Nucleon Form 
Factors as limiting cases, they clearly contain a wealth of information 
well beyond these, notably about the hitherto unrevealed orbital angular 
momenta of partons (see below). \\

The scheme presented in Fig.~\ref{fig:GPDwheel} is intended as an 
easy-to-read overview
\begin{figure}[htb]
\vspace*{0.2cm}
\hspace{3.3cm}
\includegraphics[width=12cm]{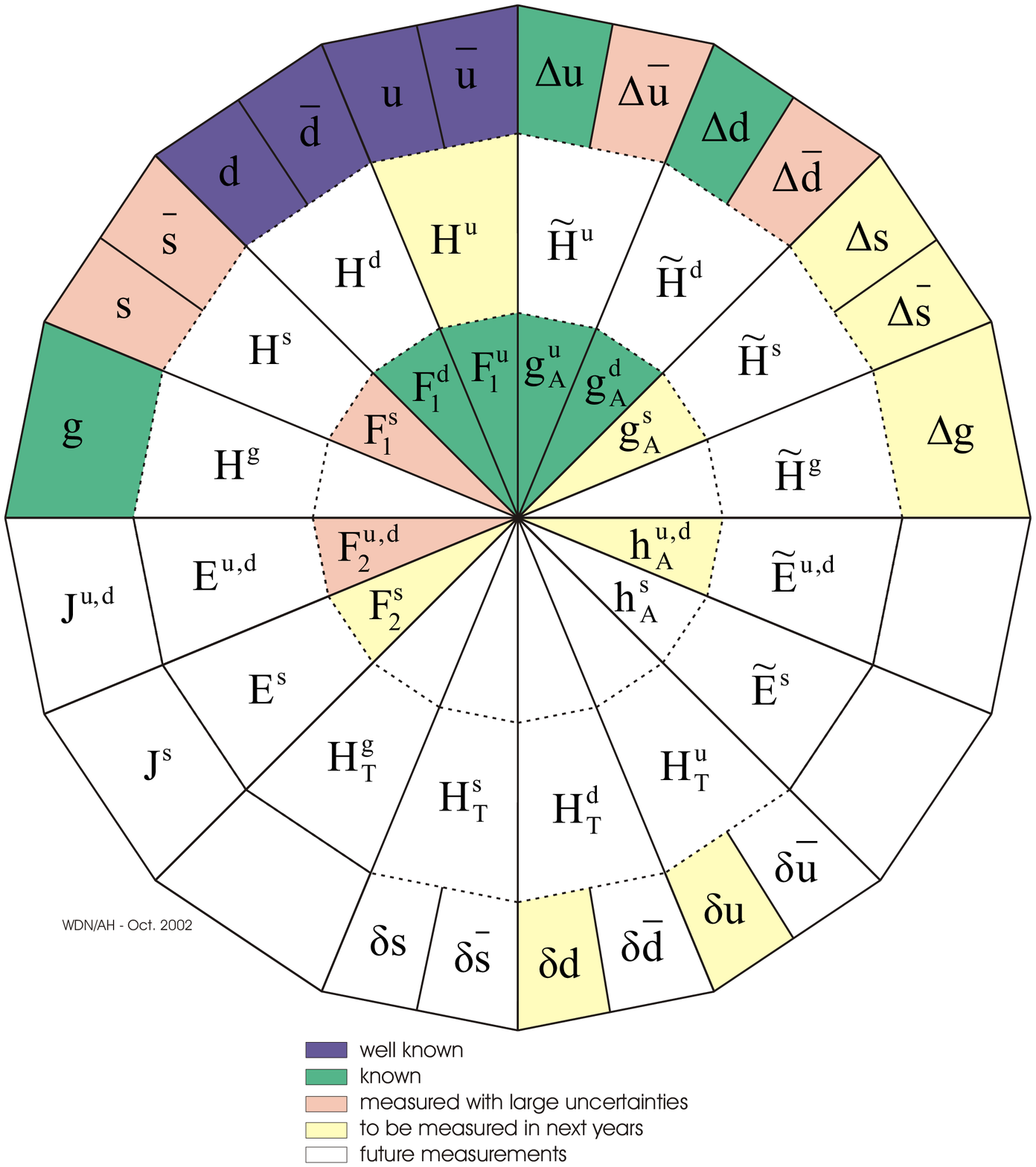}
\centering
\caption{\it Visualisation of (most of) the relevant Generalised Parton 
             Distributions and their limiting cases, forward Parton
             Distributions and Nucleon Form Factors. Different colours or
             shades of
             grey illustrate the status of their experimental access 
             (see legend). For explanations see text.}
\label{fig:GPDwheel}
\end{figure}
of the various quantities that are considered 
relevant for a comprehensive description of the angular momentum structure 
of the nucleon. It is based on the formalism of GPDs which are placed in 
the middle of three concentric rings. The two limiting cases are located 
in the adjacent rings: Nucleon Form Factors, the first moments of their 
appropriate GPDs, are shown in the innermost ring, while PDFs, their 
`forward' limits, are located in the outermost ring. Today's experimental 
knowledge of the different functions is illustrated in different colours 
or shades of grey, from light (no data exist) to dark (well known), see 
legend. The emphasis in Fig.~\ref{fig:GPDwheel} is placed 
on the physics message and not on completeness; some GPDs have been 
omitted. Empty sectors mean that the function does not exist, 
decouples from observables in the forward limit, or no strategy is known for
its measurement. GPDs will be discussed in detail in the next chapter, PDFs in 
chapter~\ref{section:PDFs}, and Nucleon Form Factors in 
chapter~\ref{section:FFs}.  \\

\section{Generalized Parton Distributions}
\begin{figure}[htb]
\noindent
\begin{minipage}[b]{.49\linewidth}
  \centering \includegraphics[width=\linewidth]{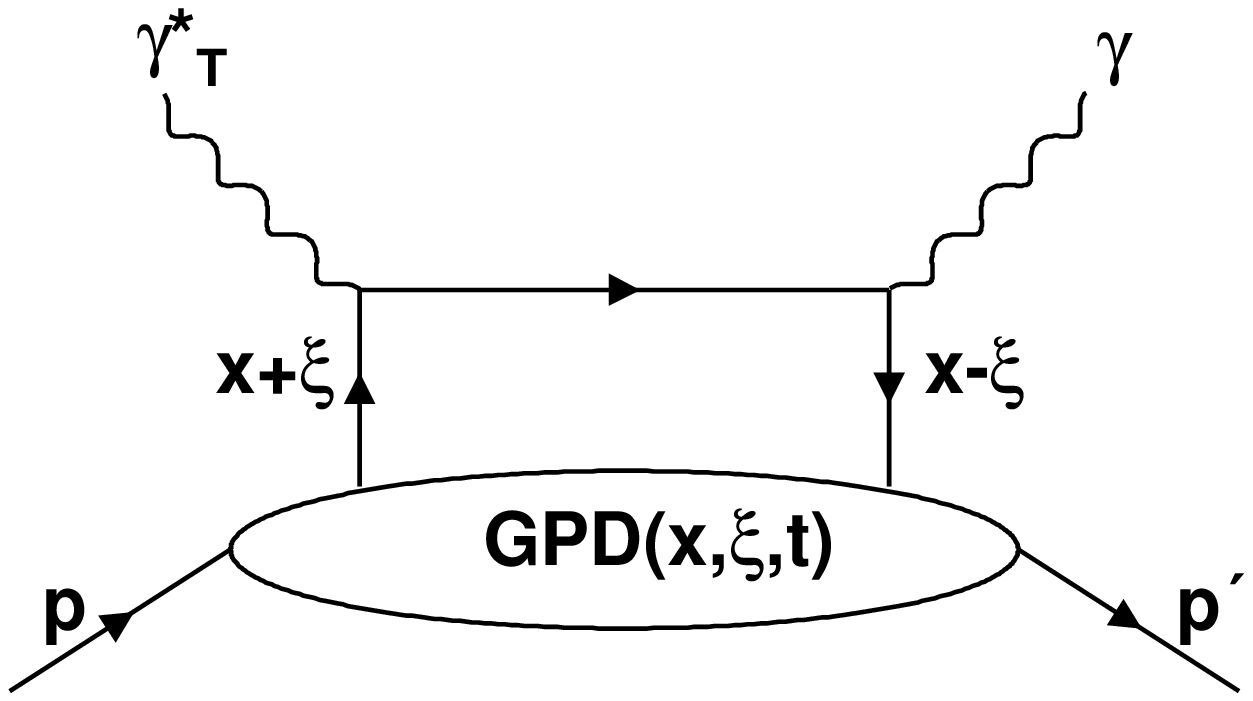}
\end{minipage} \hfill
\begin{minipage}[b]{.49\linewidth}
  \centering \includegraphics[width=\linewidth]{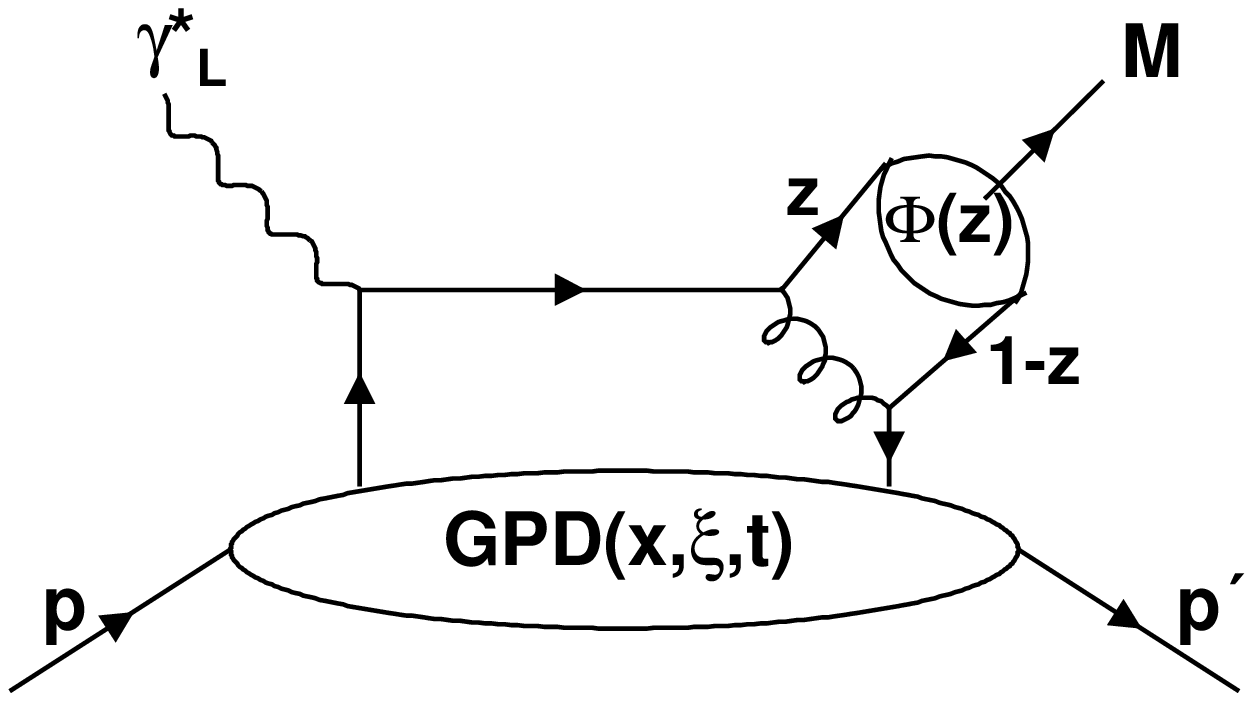}
\end{minipage}
\centering
\vspace*{-0.3cm}
\caption{\it Illustration of the two major types of hard exclusive 
             processes to extract GPDs: DVCS and Meson Production.}
\label{fig:blobs}
\end{figure}

%
Generalized Parton Distributions depend 
on a resolution scale $Q^2$, two longitudinal momentum fractions ($x, \xi$) 
and $t$, the momentum transfer at the nucleon vertex. The `skewedness' $\xi$ 
parameterises the longitudinal momentum difference of the two partons involved
in the interaction (cf. Fig.~\ref{fig:blobs}).
Through their dependence on $\xi$ the GPDs carry information about 
correlations between two different partons in the nucleon. Note that in
the `forward' limit $\xi=0$ and $t=0$.  \\

For each parton species\footnote{Here $f$ stands for the quark 
flavors $u,d,s$, but here also for the gluon index $g$.} $f$ there 
exist four GPDs ($H^f, \tilde{H}^F, E^f, \tilde{E}^f$) that do not 
flip the {\it parton} helicity and additional four 
($H_T^f, \tilde{H}_T^f, E_T^f,$ $\tilde{E}_T^f$) that do flip it 
\cite{HoodJi98,Diehl01a}.
In terms of chirality the two sets of {\it quark} GPDs can be also referred 
to as chirally-even and chirally-odd ones, respectively, while gluons 
do not have chirality. The {\it nucleon} helicity, on the other hand, is 
conserved by the $H$-functions, but not by the $E$-functions. Finally, 
the chirally-even, i.e.
parton-helicity non-flip quark GPDs can also be classified into unpolarised 
$(H^f, E^f)$ and polarised $(\tilde{H}^f, \tilde{E}^f)$ ones. The 
nucleon-helicity conserving quark GPDs $H^f$ and $\tilde{H}^f$ have as
forward limits the well-known unpolarised and longitudinally polarised quark 
PDFs $q_f, \bar{q}_f$ and  $\Delta q_f, \Delta \bar{q}_f$, respectively. 
In the case of gluons, $g$ and $\Delta g$ are the forward limits of $H^g$ 
and $\tilde{H}^g$, respectively. The nucleon-helicity non-conserving
GPDs $E^f$ and $\tilde{E}^f$ decouple from observables in the forward limit,
i.e. they have no corresponding PDFs.
As forward limit of the {\it parton-helicity flip} quark GPDs $H_T^f$ the 
transversity PDFs $\delta q_f, \delta \bar{q}_f$ are obtained, while gluons 
have no transversity~\cite{ArtruMekhfi}. \\

The recent strong interest in GPDs was stimulated by the finding~\cite{Ji} 
that the sum of the unpolarised chirally-even GPDs, $\frac{1}{2} (H^f+E^f)$, 
carries information about the {\it total} angular momentum $J^f$ of 
the parton species $f$ in the nucleon. In the limit of vanishing $t$ the 
second moment of this sum approaches $J^f$, measured at the given $Q^2$.
The total angular momentum carried by quarks of all flavors, 
$J^q = \sum_{f} J^f$, and that carried by gluons, $J^g$, are not known yet.
The same holds for the corresponding orbital momenta, $L^q$ and $L^g$. 
Through the simple integral relation $L^q = J^q - \frac{1}{2} \Delta \Sigma$,
with $\frac{1}{2} \Delta \Sigma$ being the quark's {\it spin} contribution
(cf. next chapter), a measurement of $J^q$ would allow to determine $L^q$.
Note that there is an ongoing unresolved controversy in the literature
about the appropriate definition of angular momentum operators for quarks 
and gluons~\cite{Jaffe2001}.  \\

GPDs can in principle be revealed from measurements of various cross 
sections and spin asymmetries in several exclusive processes. However,
there is no doubt that for a determination of individual GPDs from 
experimental cross sections and asymmetries a rather complicated 
procedure will have to be developed. Unlike in the case of forward PDFs 
a direct extraction of GPDs appears presently not feasible. The usual 
deconvolution procedure can not be applied, because the involved momentum 
fraction $x$ is an entirely internal variable, i.e. it is always integrated 
over in the process amplitudes. A principal way to circumvent 
this problem by distinguishing the $log \, Q^2$ behaviour from the 
$1/Q^2$ behaviour~\cite{Freund00} appears hard to realise even under
substantially improved experimental conditions in the future. Hence an 
iteration process based on the comparison of theoretical model GPDs to 
experimental data seems to be the unavoidable choice. This most probably 
will require the combination of results from various reaction channels 
into a `global fit' of the involved GPDs. Clearly, further theoretical work
on the sensitivity of experimentally available quantities to different
properties of the model functions, especially higher orders (see e.g.
Ref.~\cite{FreundMcDerm02}) and higher twists (see e.g. 
Ref.~\cite{BelMulKir2002}), seems to be of great importance.   \\

The chirally-even GPDs $(H^f, \tilde{H}^f, E^f, \tilde{E}^f)$ for $u$ and 
$d$-quarks, which are presently the most intensely discussed functions,
can be accessed experimentally in several reactions. Deeply Virtual Compton 
Scattering (DVCS), $e \, p \, \rightarrow \, e \, p \, \gamma$ (cf. left 
panel of Fig.~\ref{fig:blobs}), is presently considered the cleanest one.
In this reaction the first and, so far, the only GPD-related experimental 
results were published recently. Note that there is still no data available
that can be related to parton-helicity non-flip {\it gluon} GPDs. They can in 
principle be accessed in DVCS and meson production at small $\xi$, i.e. 
preferentially at large center-of-mass energies.   \\

A $\sin{\phi}$ behaviour\footnote{Here $\phi$ is the azimuthal angle of the 
produced real photon around the direction of the virtual photon, relative to
the lepton scattering plane.} is predicted~\cite{DiehlEtal97} for the 
azimuthal dependence of the single {\it beam-spin asymmetry} in the 
DVCS cross section. This prediction was recently proven at two different 
center-of-mass energies, by HERMES~\cite{HERMES:A_LU} and 
CLAS~\cite{CLAS:A_LU}. The asymmetry is 
given by the {\it imaginary} part of the interference term between the 
DVCS and Bethe-Heitler amplitudes. The involved combination of GPDs contains 
$H^f, \tilde{H}^f$ and $E^f$, where $H^f$ is the dominant function driven 
by kinematical factors (cf. Ref.~\cite{KN:NPA711}). Access to the
{\it real} part of the same interference term is opened by measuring the 
{\it beam-charge asymmetry} in DVCS. First results on this observable
were presented by HERMES~\cite{FE:NPA711} very recently and confirmed the 
predicted $\cos{\phi}$ behaviour. It has to be noted that the imaginary 
part is probed at the special argument $(\xi, \xi, t)$ which constitutes 
a second independent `slice' in the $(x, \xi, t)$-plane, in addition to 
the `forward slice' $(x, 0, 0)$. The $x$-dependence away from the line 
$x=\xi$ is contained in the principal value integral of the real 
part~\cite{GoePolVan2001}.  \\

\begin{figure}[h]
\centering
\vspace*{-2.3cm}
\includegraphics[width=.6\linewidth]{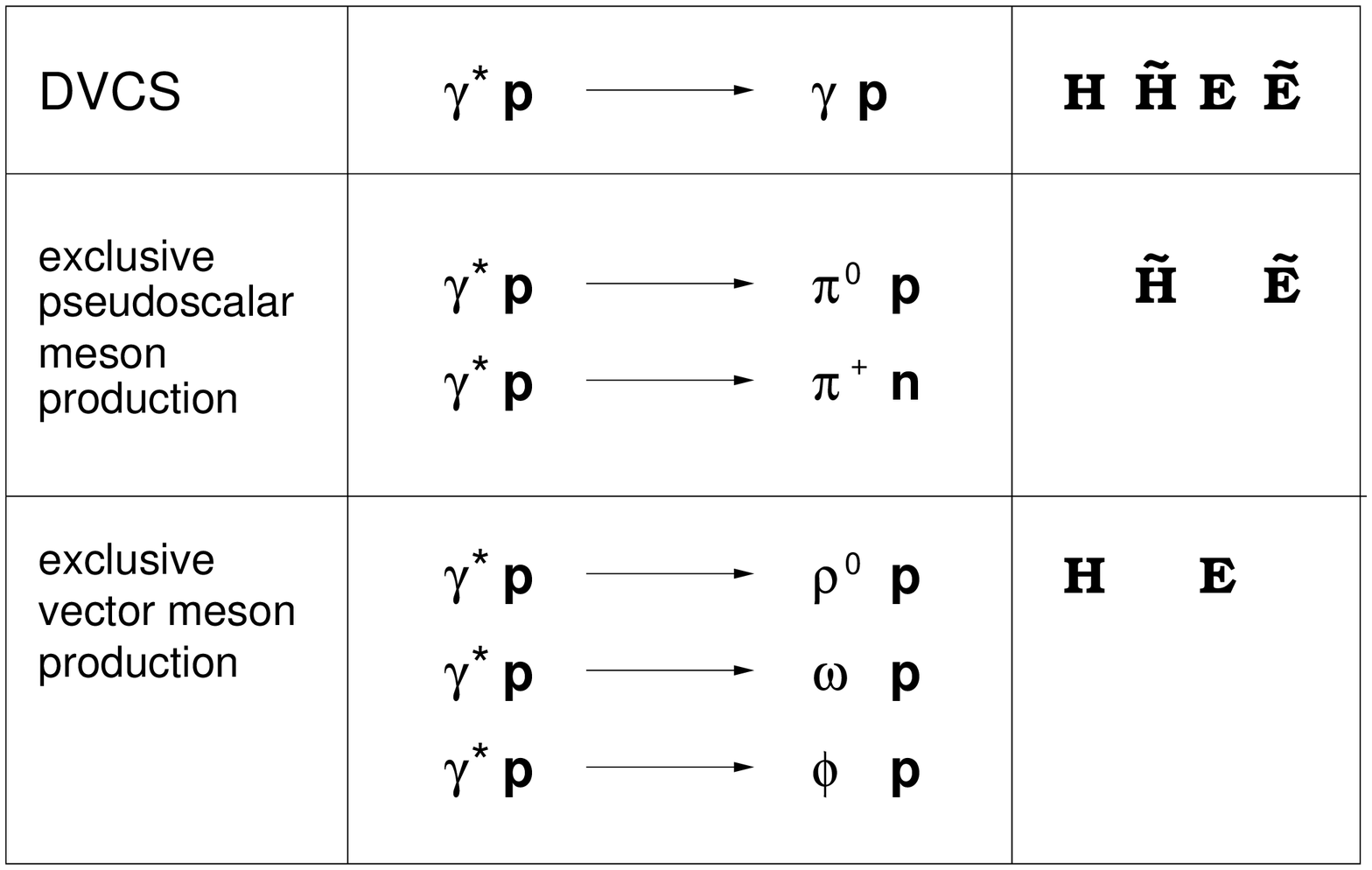}
\vspace*{-2.8cm}
\caption{\it Examples of hard exclusive processes 
                   and the involved chirally-even GPDs.}
\label{fig:DVEStable}
\end{figure}

Information on chirally-even GPDs can also be accessed in Deeply Virtual
Exclusive production of pseudoscalar and vector Mesons (DVEM), 
$e \, p \, \rightarrow \, e \, p \, M$ (cf. right panel of 
Fig.~\ref{fig:blobs}). While DVCS is suppressed in comparison 
to meson production by the additional electromagnetic coupling, the latter
is suppressed by a factor $1/Q^2$ over the former. In fixed-target
exclusive meson production, this results in an increase by a factor of 
about 10 in count rate, as compared to DVCS.
On the other hand, compared to the outgoing real photon in DVCS 
the exclusively produced meson introduces one more complication, 
namely the distribution amplitude $\Phi (z)$ as an additional unknown.
Here $z$ is the momentum fraction carried by the meson.
Cross sections and spin asymmetries for different channels are described by 
different sets of GPDs (for more details see e.g. Ref.~\cite{GoePolVan2001}). 
The comparison of pseudoscalar and vector meson production reveals that the 
final state meson acts in fact as a helicity filter. 
Fig.~\ref{fig:DVEStable} illustrates which of the chirally-even
GPDs can be accessed in which reaction channels. \\

Although there exists already a huge number of theoretical papers on 
chirally-even GPDs, only a few attempts have been made to construct models 
for them (cf. e.g. Ref.~\cite{GoePolVan2001} and references 
therein,~\cite{JiMelSong97}). They have to reproduce the $x$-dependence 
of forward PDFs and the $t$-dependence known from Nucleon
Form Factors. The crucial problem of the correlations between the different
GPD variables is still under strong theoretical debate. It remains to be
shown to which extent future experiments will be
in a position to successfully distinguish between different GPD models.
A very first glimpse of the function $H$ can be expected to 
emerge from mid-term future measurements of various hard exclusive 
reactions, especially DVCS, at the HERMES experiment upgraded with a 
Recoil Detector~\cite{HERMES:recoil} surrounding an {\it un}polarised 
target. The projected measurements with an integrated luminosity of 
2~fb$^{-1}$ will be able to distinguish between two different sets of 
`contemporary' model GPDs over a certain range in the longitudinal 
momentum variable $\xi = x_B/(2-x_B)$, where $x_B$ is the Bjorken scaling 
variable. (For details see Refs.~\cite{KN:NPA711,KN:EPJC23}).   \\

Considerably less theoretical work and no experimental results exist on
either quark or gluon {\it helicity-flip} GPDs ($H_T^f, \tilde{H}_T^f, 
E_T^f, \tilde{E}_T^f$). The first proposal how to 
experimentally access helicity-flip {\it quark} GPDs is based on the
simultaneous production of two $\rho$ mesons~\cite{IvanPire02} at large
rapidity separation. This may experimentally be feasible but requires 
large center-of-mass energies. No model has been proposed yet for these 
functions and therefore no projections are available yet. Helicity-flip 
{\it gluon} GPDs can in principle be accessed through measurements of a 
distinct angular dependence of the DVCS cross 
sections~\cite{HoodJi98,BelMuel2000}.  \\

As mentioned above, the forward limit of the second moment of the sum
$\frac{1}{2} (H^f + E^f)$ determines the total angular momentum $J^f$ of 
the parton species $f$ in the nucleon. In this context a forward {\it orbital 
angular momentum (OAM) distribution} was introduced~\cite{HoodJiLu99} as 
$L^f(x) = J^f(x) - \Delta q_f(x)$, where $J^f(x)$ is the forward limit of 
the GPD sum $\frac{1}{2} (H^f + E^f)$ and $\Delta q_f(x)$ is the helicity
distribution (cf. next chapter). Since the forward limit of the GPD $H^f$ 
is known from the well-measured PDFs $q_f$ and $\bar{q}_f$, a determination 
of $J^f(x)$ requires a measurement of the GPD $E^f$. This is expressed
in Fig.~\ref{fig:GPDwheel} by placing the symbol $J^f$ in the outermost ring
of the $E^f$ sector. Experimental access to $E^f$ may be achieved through 
DVCS measurements with an unpolarised beam and a {\it transversely} polarised 
target. Projections for the statistical accuracy attainable when measuring 
the relevant single target-spin asymmetry $A_{UT}(\phi)$ are given in 
Ref.~\cite{wdn:Trento} for both a possible low-luminosity (0.8 fb$^{-1}$) 
measurement in the mid-term future and a high-luminosity measurement 
(100 fb$^{-1}$) in the far future.
Note that the present HERMES measurements using a transversely polarised 
target, aiming at a determination of $u$ and $d$ quark transversity (cf. 
next chapter), are planned to collect about 0.15 fb$^{-1}$. Also, important
additional knowledge on $E^f$ can be expected from exclusive vector meson
production with transverse target polarisation~\cite{GoePolVan2001}.  \\

\section{Parton Distribution Functions}
\label{section:PDFs}
A Parton Distribution as forward limit of a Generalised Parton 
Distribution depends only on the constituent's momentum fraction $x$ 
that in deep inelastic lepton-nucleon scattering (DIS) is identified with 
the Bjorken scaling variable $x_B$. Additionally, it depends on a 
resolution scale $Q^2$, as mentioned above. On the simplest (twist-2) 
level, as long as multi-parton correlations are not considered, the 
complete set of quark PDFs consists (for every flavor $u,d,s$) of number 
density $q_f$, {\it longitudinally} polarised (helicity) distribution 
$\Delta q_f$, and {\it transversity} distribution $\delta q_f$. Only these 
three functions together form the minimum data set necessary for meaningful 
confrontations with theoretical models of the ground state of the nucleon.  \\

In-depth studies of DIS over the last two decades resulted in unpolarised 
quark distributions that are precisely known for valence and well known 
for sea quarks~\cite{PDG}. Their longitudinally polarised counterparts 
were measured only recently~\cite{HERMES:Delta-q}, still with large
uncertainties for the sea distributions.
In contrast, the {\it transverse} spin structure of the nucleon is still 
completely unexplored, even in the forward limit. Several experiments will 
deliver first data in the near future: COMPASS at CERN~\cite{COMPASS:prop}, 
HERMES at DESY~\cite{KNO} and RHIC-spin at BNL~\cite{RHIC-spin:summary}.  
Its interpretation will to a large extent be based on the analysis of 
(single) spin asymmetries in (semi-inclusive) hadron production cross 
sections. However, precise information on sea quark as well as highly 
precise data on valence quark distributions can only be expected from 
the next generation of electron-nucleon scattering experiments. For a 
brief overview including projections and further references see, e.g., 
Ref.~\cite{wdn:Trento}. \\

For a given flavor, the first moment of the sum (difference) of quark
and anti-quark helicity (transversity) distributions represents the axial 
(tensor) charge of the nucleon. Precise measurements of these integrals 
are of special interest. The flavor sum of the axial charges, 
$\Delta \Sigma (Q^2)$, describes the quark's contribution to the total 
longitudinal spin of the nucleon. Experimentally, a value far below the 
originally expected non-relativistic limit is found, irrespectively 
of the underlying renormalisation scheme. This behaviour gave rise to 
the `spin crisis' of the nineties. It may be attributed to the fact that
the $Q^2$-evolution of $\Delta \Sigma$ involves the polarised gluon
distribution. In contrast, the hitherto totally unmeasured flavor sum of 
the tensor charges, $\delta \Sigma (Q^2)$, decouples as an all-valence object 
from gluons and sea quarks and hence is expected to be much closer to 
the non-relativistic limit (cf. the discussion on lattice QCD results 
below). Measurements of the tensor charge will 
give access to the hitherto unmeasured chirally-odd operators in QCD which
are of great importance to understand the role of chiral symmetry in the 
structure of the nucleon~\cite{Jaffe97}. 
Also, the tensor charge is required as input 
to calculate the electric dipole moment of the neutron~\cite{Ji:SPIN2002}
in beyond-the-standard-model theories where the quarks may have electrical
dipole moments themselves~\cite{DemirEtal02}.   \\

Lattice calculations, performed in the context of the operator-product 
expansion (OPE), lead to reliable results on the `valence' tensor charge 
$\delta \Sigma \approx \delta u + \delta d$; their precision improves 
with time as better methods and computers are used (e.g. $0.562 \pm 0.088$ 
in 1997~\cite{Aoki97}, $0.746 \pm 0.047$ in 1999~\cite{Capitani99}). 
Nevertheless, these values are far away from the non-relativistic limit of 
$\frac{5}{3}$ expected for a nucleon consisting of just three valence quarks. 
In contrast, lattice OPE calculations do not yet lead to reliable results 
on the `valence' axial charge 
$\Delta \Sigma \approx \Delta u + \Delta d$, because the quark-line 
disconnected diagram cannot be calculated yet. The `quenched' 
approximation leads to the value 
$\Delta \Sigma = 0.18 \pm 0.10$~\cite{Aoki97}. This
is surprisingly consistent with the earlier result $0.18 \pm 0.02$ obtained
with the simpler method of dynamically staggered fermions~\cite{Altmeyer94}.
Note that next-to-leading order QCD analyses of experimental data yield
$\Delta \Sigma = 0.2-0.4$, depending on the factorisation 
scheme~\cite{FiliJi2001}. All
numbers quoted above refer to $Q^2$-values of a few GeV$^2$.  \\

\section{Elastic Form Factors}
\label{section:FFs}
The $t$-dependent elastic form factors of the nucleon can in principle be 
obtained from the Generalised Parton Distributions (once they are known) by
integrating over their momentum-fraction variables, i.e. as their first 
moments. The first moments of the unpolarised chirally-even GPDs $H^f$ and 
$E^f$ constitute the Dirac (charge) and Pauli (current) form factors 
$F_1^f(t)$ and $F_2^f(t)$. Note that measurements are usually done for the 
electric and magnetic Sachs form factors $G_E$ and $G_M$ which are linear 
combinations of the Dirac and Pauli form factors. The polarised 
chirally-even GPDs $\tilde{H}^f$ and $\tilde{E}^f$ yield the axial-vector 
and pseudoscalar form factors $g_A^f$ and $h_A^f$ (also denoted as $G_A$ 
and $G_P$). Also for the helicity-flip quark GPDs ($H_T^f, \tilde{H}_T^f, 
E_T^f, \tilde{E}_T^f$) form factors exist; for example the first moment of
$H_T^f$ at $t=0$ is the
tensor charge of quark species $f$. The corresponding sectors in 
Fig.~\ref{fig:GPDwheel} contain no symbols because in the literature there 
is no unique naming convention nor are there ways for their measurement. For
that reason the gluon form factors are also omitted.  \\

Measurements of proton and neutron elastic form factors are 
accomplished by widely different experimental approaches. Beam energies 
of a few GeV and less are sufficient to determine their $Q^2$-dependence 
which can, in most cases, be approximately described by the $1/Q^2$ dipole 
form factor. For a recent review of existing measurements and for further 
references, see Ref.~\cite{KellyYer2002}.  \\

\section{Experimental Requirements}
\label{section:exp}
Presently, almost the entire field of GPD-related physics constitutes 
experimentally virgin territory. Stimulated
by the physics insight that can be expected from measurements
of GPDs, experimentalists are striving to optimise and upgrade their 
existing facilities to obtain experimental results that can be used to 
extract first information about GPDs themselves.  \\

There is little doubt that a completely new step in experimentation,
concerning both fixed-target and collider facilities, is required for 
future GPD measurements. In Fig.~\ref{fig:PipkeTable} 
a brief summary of the abovementioned major physics topics is given
in conjunction with experimental requirements for a future 
fixed-target electron-nucleon scattering facility. It must offer luminosities
of at least 10$^{35}$ per cm$^2$s requiring an accelerator with a duty cycle
of 10\% or more. Beam energies above 30 GeV are needed to cover a kinematic
range suitable for extracting cross sections and their scale dependence in
exclusive measurements. For the non-exclusive studies of the hadron structure,
including precise measurements of polarised Parton Distribution Functions,
variable beam energies in the range of 50-200 GeV are required. In both cases,
highest possible polarisation of beam and targets as well as the availability 
of both beam charges is required
to fully exploit the potential of asymmetry measurements. Large-acceptance 
detector systems with high-rate capabilities must reach a mass resolution of 
a third of the pion mass, mandatory for the measurement of exclusive channels. 
For more detailed discussions of the various present-day options for such a 
facility see, e.g., Refs.~\cite{wdn:Trento,rk:SPIN2002} which also contain
further references. Note that precise measurements of 
Nucleon Form Factors, while constituting an essential part of the physics 
menu addressed in this paper, will be pursued at lower-energy facilities.  \\

\section{Conclusions}
As can be concluded from today's knowledge, major steps in both theory and 
experimentation are required to accomplish a comprehensive understanding of 
the angular momentum structure of the nucleon. Physics objectives can be 
identified for a future fixed-target electron-nucleon scattering facility 
with polarised targets, high-luminosity, and variable beam energies of 
30-200 GeV. Precise measurements of hard exclusive processes will have to 
be performed in several reaction channels towards a determination of as many
as possible Generalised Parton Distributions in a `global' fit. Independently,
very precise measurements of the GPD limiting cases, namely forward Parton 
Distribution Functions on the one hand, and Nucleon Form Factors on the 
other, will be of great importance for further developments of the theory, 
because they eventually will have to complement and confirm the GPD results.  \\

\section*{Acknowledgements}
\begin{figure}[h]
\hspace{3.3cm}
\includegraphics[width=11.5cm]{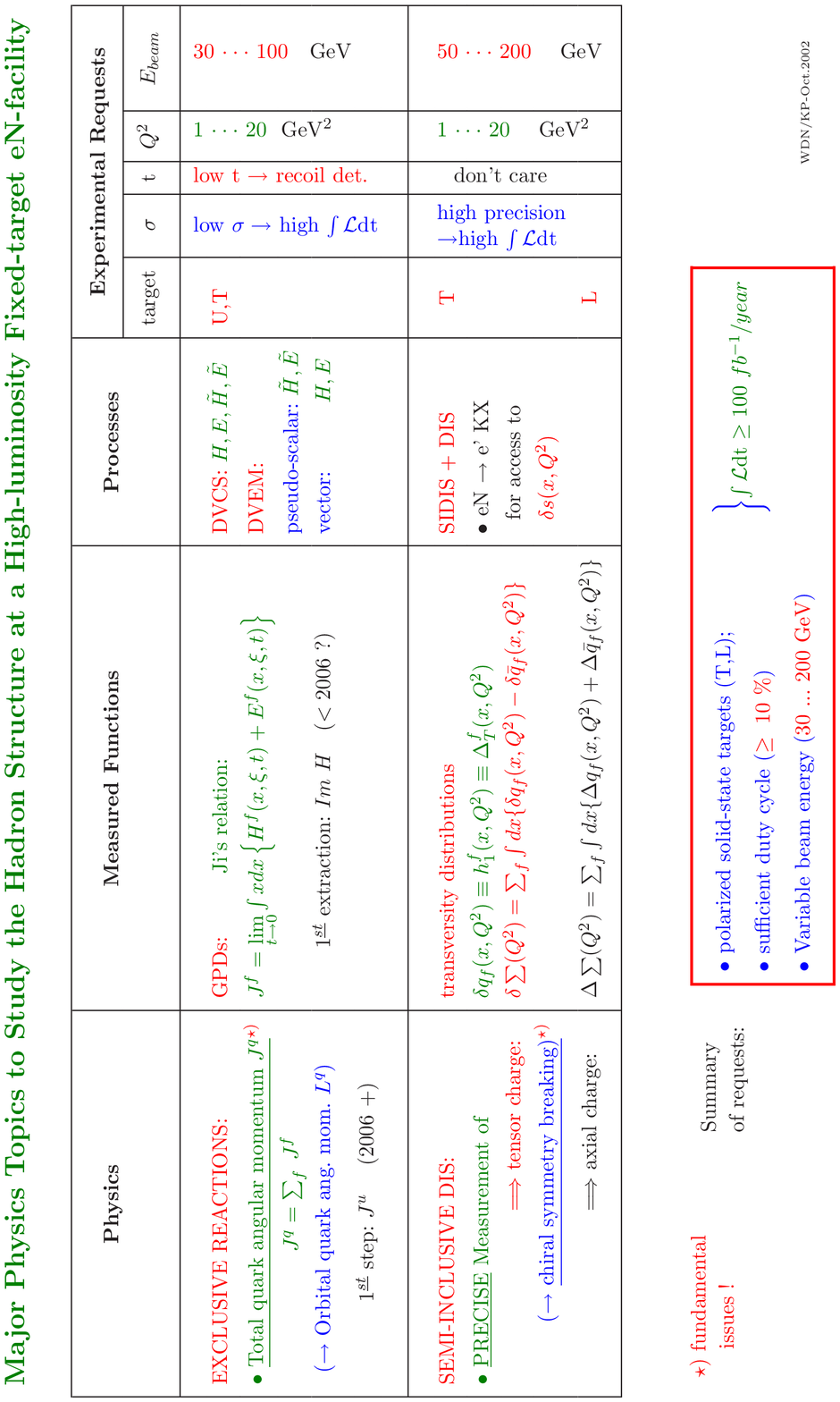}
\centering
\vspace*{0.4cm}
\caption{\it Table of major physics topics in conjunction with
             experimental requirements.}
\label{fig:PipkeTable}
\end{figure}

%
I am deeply indebted to M.~Diehl and X.~Ji for enlightening discussions, 
to F.~Ellinghaus, R.~Kaiser and J.~Volmer for a careful reading 
of the manuscript, and to K.~Jansen and G.~Schierholz for valuable
comments on the lattice QCD section. Many thanks to B.~Seitz for 
commenting on, and to A.~Hagedorn for producing Fig.~\ref{fig:GPDwheel}.
The help of K.~Pipke in preparing Fig.~\ref{fig:PipkeTable} is
acknowledged.


\newpage

\begin{thebibliography}{99}

\bibitem{Mueller} D.~M\"{u}ller {\it et al.}, Fortschr. Phys. {\bf 42},
                  101 (1994).

\bibitem{Ji} X.~Ji, Phys. Rev. Lett. {\bf 78}, 610 (1997); 
                    Phys. Rev. {\bf D 55}, 7114 (1997).

\bibitem{Radyushkin} A.V.~Radyushkin, Phys. Lett. {\bf B 380}, 417 (1996);
                     Phys. Rev. {\bf D 56}, 5524 (1997).

\bibitem{Bluemlein} J. Bl\"umlein {\it et al.}, Nucl. Phys. {\bf B 560}, 
                    283 (1999); Nucl. Phys. {\bf B 581}, 449 (2000).

\bibitem{HoodJi98} P.~Hoodbhoy, X.~Ji, Phys. Rev. {\bf D 58}, 054006 (1998).

\bibitem{Diehl01a} M.~Diehl, Eur. Phys. J. {\bf C 19} 485 (2001).

\bibitem{ArtruMekhfi} X.~Artru, M.~Mekhfi, Z. Phys. {\bf C 45} 669 (1990).

\bibitem{Jaffe2001} R.L.~Jaffe, Phil. Trans. Roy. Soc. Lond. {\bf A 359},
                    391 (2001) [arXiv:hep-ph/0008038].

\bibitem{Freund00} A.~Freund, Phys. Lett. {\bf B 472}, 412 (2000).

\bibitem{FreundMcDerm02} A.~Freund, M.~McDermott, Eur. Phys. J. {\bf C 23},
                         651 (2002).

\bibitem{BelMulKir2002} A.V.~Belitsky, D.~M\"{u}ller, A.~Kirchner,
                        Nucl. Phys. {\bf B 629}, 323 (2002).

\bibitem{DiehlEtal97} M.~Diehl {\it et al.}, Phys. Lett {\bf B 411}, 193
                      (1997).

%
\bibitem{HERMES:A_LU} HERMES coll., A.~Airapetian {\it et al.}, 
                      Phys. Rev. Lett. {\bf 87}, 182001 (2001).

\bibitem{CLAS:A_LU} CLAS coll., S.~Stepanyan {\it et al.},
                      Phys. Rev. Lett. {\bf 87}, 182002 (2001).

\bibitem{KN:NPA711} V.A.~Korotkov, W.-D.~Nowak, Nucl. Phys. {\bf A 711}, 
                    175 (2002) [arXiv:hep-ph/0207103].

\bibitem{FE:NPA711} F.~Ellinghaus for the HERMES coll., Nucl. Phys. 
                    {\bf A 711}, 171 (2002) [arXiv:hep-ex/0207029].

\bibitem{GoePolVan2001} K.~Goeke, M.V.~Polyakov, M.~Vanderhaeghen,
                        Progr. Part. Nucl. Phys. {\bf 47}, 401 (2001).

\bibitem{JiMelSong97} X.~Ji, W.~Melnitchouk, X.~Song, Phys. Rev. 
                      {\bf D 56}, 5511 (1997).

\bibitem{HERMES:recoil} HERMES coll., DESY PRC 01-01, 2002.

\bibitem{KN:EPJC23} V.A.~Korotkov, W.-D.~Nowak, 
                    Eur. Phys. J. {\bf C 23}, 455 (2002).

\bibitem{IvanPire02} D.Yu.~Ivanov {\it et al.}, CPHT RR 059.0602, 
                     arXiv:hep-ph/0209300.

\bibitem{BelMuel2000} A.V.~Belitsky, D.~M\"{u}ller, Phys. Lett. 
                      {\bf B 486}, 369 (2000).

\bibitem{HoodJiLu99} P.~Hoodbhoy, X.~Ji, W.~Lu, Phys. Rev. {\bf D 59}, 
                     014013 (1999). 

\bibitem{PDG} Particle Data Group, D.E.~Groom {\it et al.}, 
              Eur. Phys. J. {\bf C 15}, 1 (2000).

\bibitem{HERMES:Delta-q} M.~Beckmann for the HERMES coll., 
                         arXiv:hep-ex/0210049, to be publ. in Proc. of the 
         `Workshop on Testing QCD through Spin Observables in Nuclear 
          Targets', Charlottesville, Virginia/USA, Apr. 18-20, 2002.

\bibitem{COMPASS:prop} COMPASS coll., CERN/SPSLC 96-14 (1996).

\bibitem{KNO} V.A.~Korotkov, W.-D.~Nowak, K.~Oganessyan,
              Eur. Phys. J. {\bf C 18}, 639 (2001).

\bibitem{RHIC-spin:summary} G.~Bunce {\it et al.}, Ann. Rev. Nucl. 
        Part. Sci. {\bf 50}, 525 (2000).

\bibitem{wdn:Trento} W.-D.~Nowak, Nucl. Phys. {\bf B} (Proc. Suppl.) 
        {\bf 105}, 171 (2002).

\bibitem{Jaffe97} R.L.~Jaffe, MIT-CTP-2685, arXiv:hep-ph/9710465.

\bibitem{Ji:SPIN2002} X.~Ji, Proc. of the `15th Int. Spin Symposium', 
                      Long Island, New York/USA, Sept. 9-14, 2002.

\bibitem{DemirEtal02} D.A.~Demir, M.~Pospelov, A.~Ritz, arXiv:hep-ph/0208257.

\bibitem{Aoki97} S.~Aoki {\it et al.}, Phys. Rev. {\bf D 56}, 433 (1997).

\bibitem{Capitani99} S.~Capitani {\it et al.}, Nucl. Phys. {\bf B}
                     (Proc. Suppl.) {\bf 79}, 548 (1999). 

\bibitem{Altmeyer94} R.~Altmeyer {\it et al.}, Phys. Rev. {\bf D 49},
                     3087 (1994).

\bibitem{FiliJi2001} B.W.~Filippone, X.~Ji, arXiv:hep-ph/0101224.

\bibitem{KellyYer2002} J.J.~Kelly, these Proceedings and Proc. of the 
                      `15th Int. Spin Symposium', 
                      Long Island, New York/USA, Sept. 9-14, 2002.

%
\bibitem{rk:SPIN2002} R.~Kaiser, Proc. of the `15th Int. Spin Symposium', 
                      Long Island, New York/USA, Sept. 9-14, 2002.


\end{thebibliography}
\end{document}